\documentclass[11pt]{article}
\usepackage[a4paper, portrait, margin=2.5cm]{geometry}
\usepackage{graphicx}
\usepackage{lastpage}
\usepackage{fancyhdr}
\usepackage{authblk}
\usepackage{lineno}
\usepackage{amsmath}

\fancypagestyle{firstpage}{%
  \setlength\headheight{154pt}
  \fancyhf{}%
  \cfoot{\footnotesize Page \thepage\ of \pageref{LastPage}}
}

\begin{document}
\title{Measurement of diboson production and precision EFT constraints in ATLAS} 
\date{Presented at the 32nd International Symposium on Lepton Photon Interactions at High Energies, Madison, Wisconsin, USA, August 25-29, 2025}
\author[a,b]{Shu Li}
\author[ ]{\textit{On behalf of ATLAS Collaboration}}

\affil[a]{State Key Laboratory of Dark Matter Physics, Tsung-Dao Lee Institute, Shanghai Jiao Tong University, 1 Lisuo Road, Shanghai 201210, China}
\affil[b]{State Key Laboratory of Dark Matter Physics, Institute of Nuclear and Particle Physics, School of Physics and Astronomy, Key Laboratory for Particle Astrophysics and Cosmology (MOE), Shanghai Key Laboratory for Particle Physics and Cosmology (SKLPPC), Shanghai Jiao Tong University, 800 Dongchuan Road, Shanghai 200240, China}

\renewcommand\Authand{, }

\newgeometry{top=2cm, bottom=7cm}
\maketitle
\thispagestyle{firstpage}
\abstract{
These proceedings summarize the latest progress  by the ATLAS Experiment at the LHC in measuring diboson production and related searches for physics beyond Standard Model via anomalous gauge couplings with the latest Effective Field Theory approach. The most recent measurements of $W^{+}W^{-} \to \ell^{+}\nu\ell^{-}\bar{\nu}$, $W^{\pm}Z \to 3\ell1\nu$ and $Z(\to \ell^{+}\ell^{-})\gamma$ measurements with ATLAS full Run 2 dataset are presented, along with the highlights of the first evidence of $W^{+}W^{-}$ charge asymmetry, the first measurement of CP-violation sensitive observables in $WZ$, and, for the first time at the LHC, $SU(2)_L \otimes U(1)_Y$ fully gauge invariant anomalous neutral triple gauge coupling limits with $Z(\to \ell^{+}\ell^{-})\gamma$ process.
}
\restoregeometry

\begin{figure}[htb]
\centering
\includegraphics[width=0.95\textwidth]{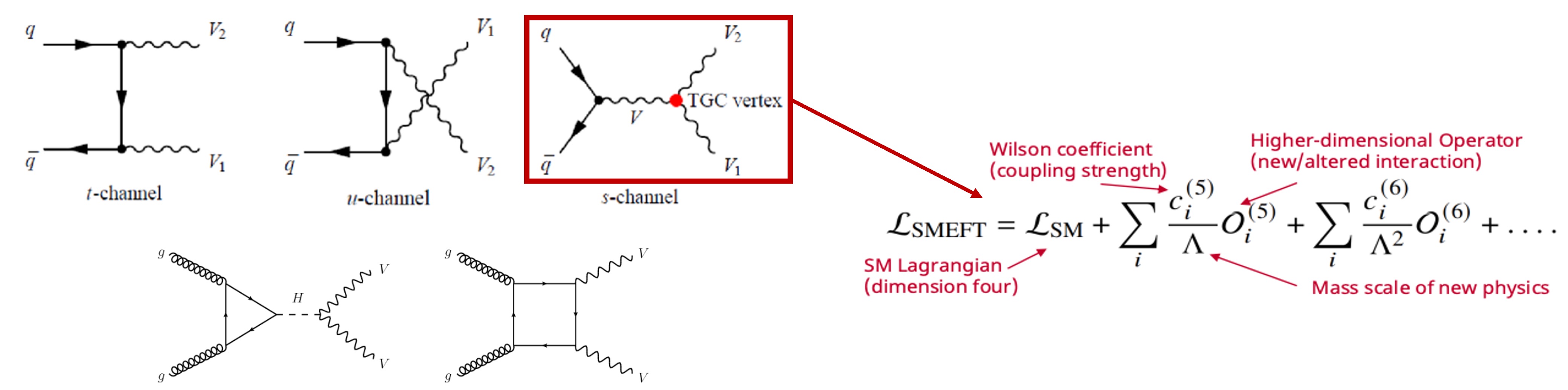}
\caption{SM EWK diboson production diagrams and the correspondance of SM EFT lagrangian with high dimension operators inducing aTGCs.}
\label{fig:diboson-diag}
\end{figure}

The Standard Model (SM) Lagrangian describes the most elementary interactions in the miscroscopic universe unifying the electromagnetic, weak and strong forces within its theory framework
along with the building blocks of matter, the fermions. Bosons, being the elementary force carriers, play pivotal roles in releasing the foundamental interactions and the corresponding
symmetries, the breaking of the symmetries, and the kinematics behind their interactions. Given the importance of understanding the foundamental interactions, SM electroweak (EWK) boson physics has been
one of key research domains in energy frontier collider physics, in particular the SM EWK diboson physics. SM EWK diboson physics (Feynman Diagrams in Figure~\ref{fig:diboson-diag}),
besides its natural role of testing EWK sector with high precision measurement results in comparison with perturbative QCD and EWK high order theory predictions,
provides also solid understanding of irreducible backgrounds as pivotal foundation for Higgs to diboson decay measurements, SM EWK measurements of rare processes, including
Vector Boson Scattering (VBS), triboson production, and related Beyond SM (BSM) searches including anomalous triple gauge coupling (aTGC) searches with SM Effective Field Theory (EFT)
high dimension operators as shown in Figure~\ref{fig:diboson-diag}.

\begin{figure}[htb]
\centering
\includegraphics[width=0.5\textwidth]{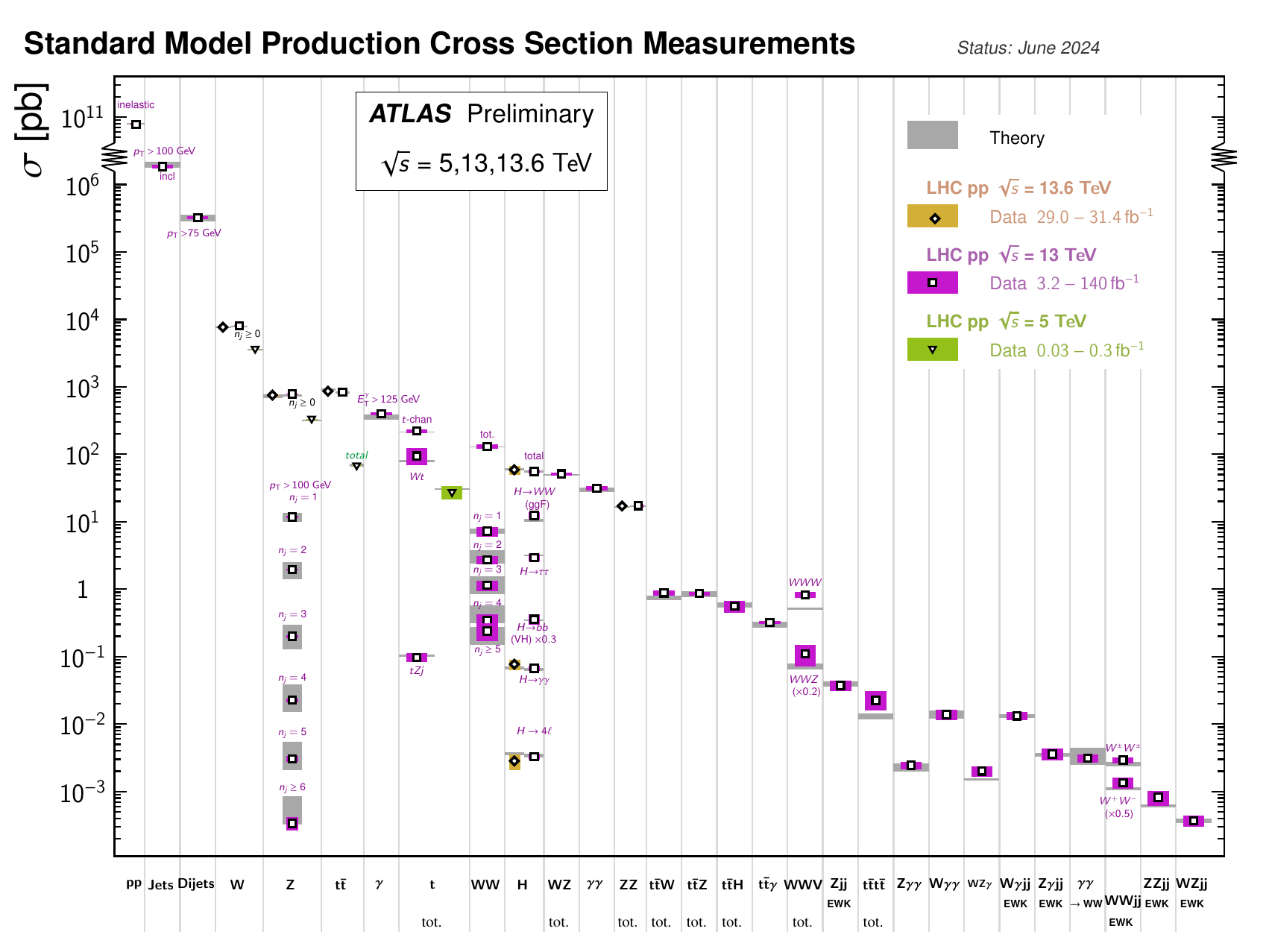}
\includegraphics[width=0.45\textwidth]{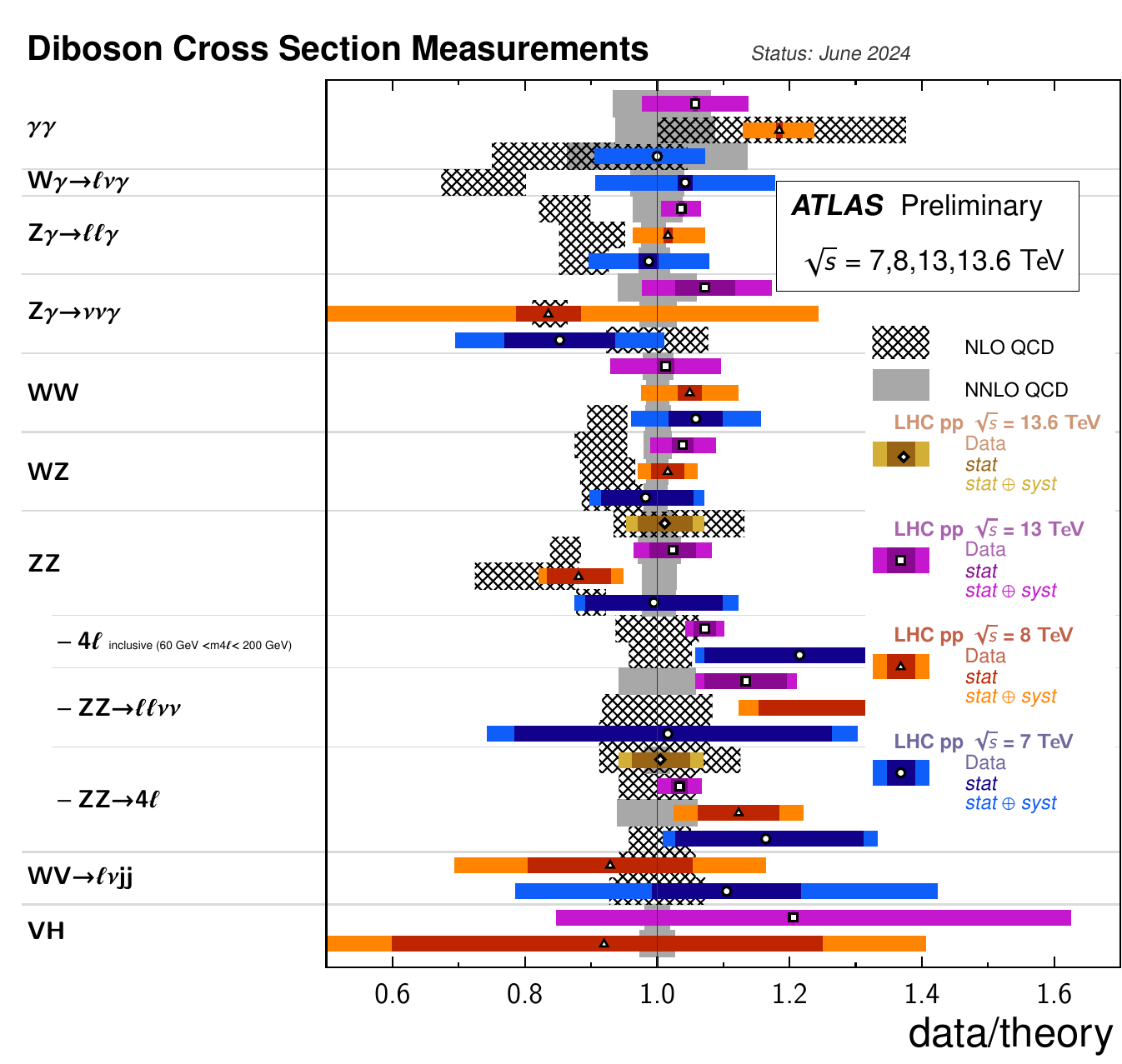}
\caption{The latest SM measurement summary plot (Left) and diboson measurement summary plot (Right)~\cite{bib:SM-Summary}.}
\label{fig:SM-Summary}
\end{figure}

ATLAS~\cite{bib:ATLAS-Det}, being one of two general purpose experiements at the Large Hadron Collider (LHC), has produced fruitful EWK diboson physics results throughout Run 1 and Run 2.
While ATLAS Run3 proton{\textendash}proton collider data-taking is delivering more data statistics in the pipeline smoothly at 13.6 TeV center-of-mass energy, the SM EWK physics success continues.
In one of the cleanest signatures, $ZZ\to 4\ell$ decay processes, the first ATLAS Run3 diboson measurement was accomplished and published last year in 2024~\cite{bib:ZZ-Run3} and presented
at many former international conferences. Despite more precision diboson channel Run3 results still take time to finish, new ATLAS full Run 2 results with unprecedented precisions in diboson
processes still drive the new milestones in energy frontier EWK physics with selective highlights presented in this talk. The latest SM summary of measured total and fiducial cross sections
are shown in Figure~\ref{fig:SM-Summary} along with deciated summary of diboson cross sections. More complete and detailed summaries can be found in Ref.~\cite{bib:SM-Summary}.

\begin{figure}[htb]
\centering
\includegraphics[width=0.4\textwidth]{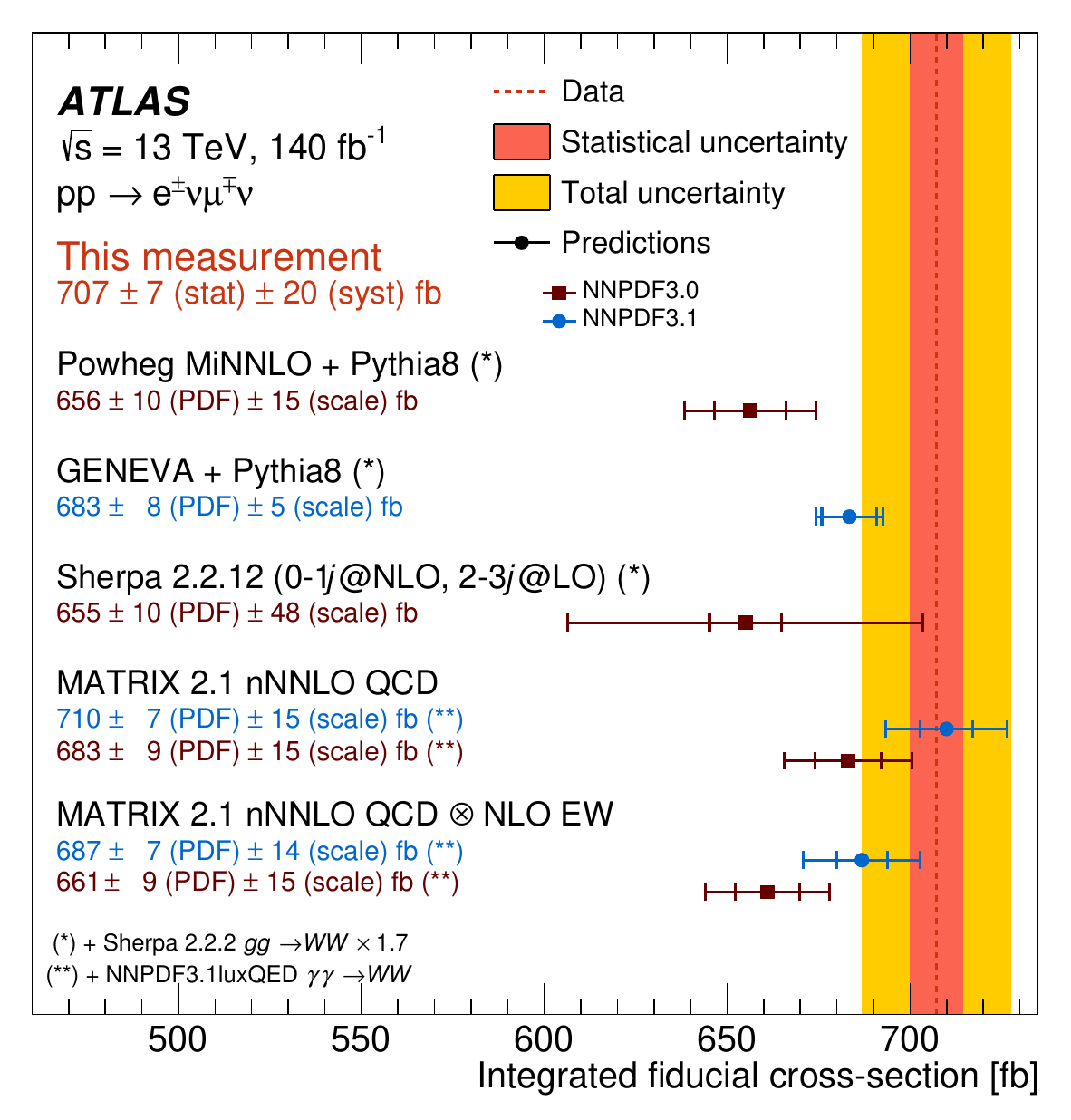}
\includegraphics[width=0.55\textwidth]{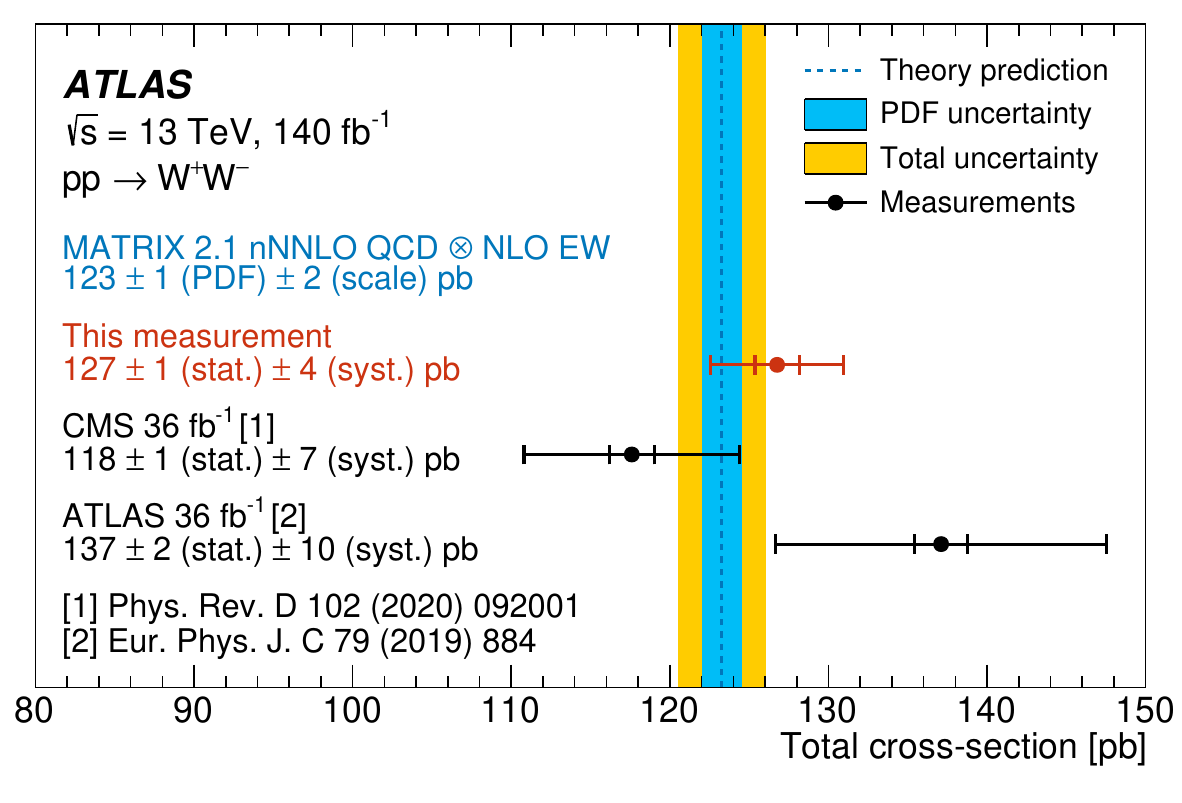}
\caption{The measured fiducial and total cross sections of $pp\to WW$ with $e\mu$ decay channel~\cite{bib:WW-FullRun2}.}
\label{fig:WW-xsec}
\end{figure}

In the fully leptonic decay channel of diboson $W^{+}W^{-}$ productions, ATLAS full Run 2 analysis~\cite{bib:WW-FullRun2}, with 140 fb$^{-1}$ $pp$ collision data, performed the total,
fiducial and differential cross secton measurements in the different flavor and opposite charge $e^{\pm}\mu^{\mp}$ channel so as to suppress the enormous background contaminations from
Drell-Yan processes. Twelve physical observables are measured with differental cross sections including $p_{\text{T}}^{\text{lead\:lep}}$, $p_{\text{T}}^{\text{sub-lead\:lep}}$, $p_{\text{T}}^{e\mu}$, $y_{\text{T}}^{e\mu}$, $m_{e\mu}$,
$\Delta \phi_{e\mu}$, $\text{cos}\theta^{\ast} = |\text{tanh}(\Delta\eta(e,\mu)/2)|$, $E_{\text{T}}^{\text{miss}}$, $m_{\text{T},e\mu}$, $H_{\text{T}}^{\text{lep+MET}}$, $S_{\text{T}}$, $N_{\text{jets}}$.
In the jet-inclusive phase space, the measured precision has been for the first time reached $\sim$3.1\%.
The measured fiducial and total cross sections in Figure~\ref{fig:WW-xsec} are consistent with by far the most precise theory predictions:
nNNLO QCD $\otimes$ NLO EWK using latest NNPDF3.1 luxQED PDF, which gives better agreement compared with NNPDF3.0 PDF. A new highlight in this analysis is that Charge asymmetry $A_C$ is
measured as a function of absolute difference of leptons absolute rapidities $||\eta_{\ell^{+}}| - |\eta_{\ell^{-}}||$, and invariant mass of the dilepton system, in comparison with NNLO+PS
predictions from Powheg MiNNLO+Pythia8. The measured result shows significant asymmetry, which is consistent with the theory anticipation given the 2:1 ratio of up:down valence quarks.
Moreover, $m_{\text{T},e\mu}$ differential distribution is used for aTGC limit setting with seven SM EFT Wilson coefficients in Warsaw basis.

\begin{figure}[htb]
\centering
\includegraphics[width=0.45\textwidth]{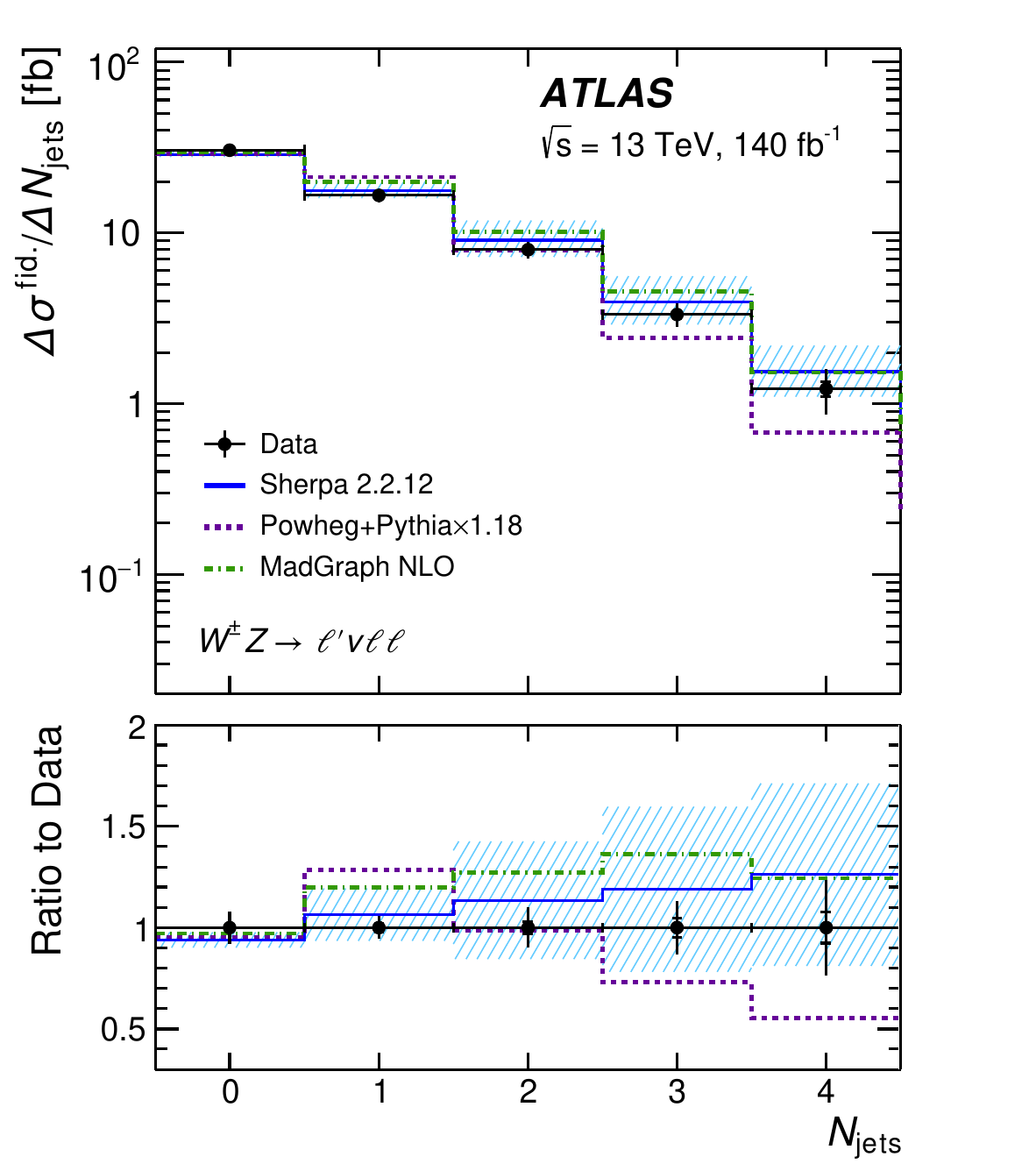}
\includegraphics[width=0.45\textwidth]{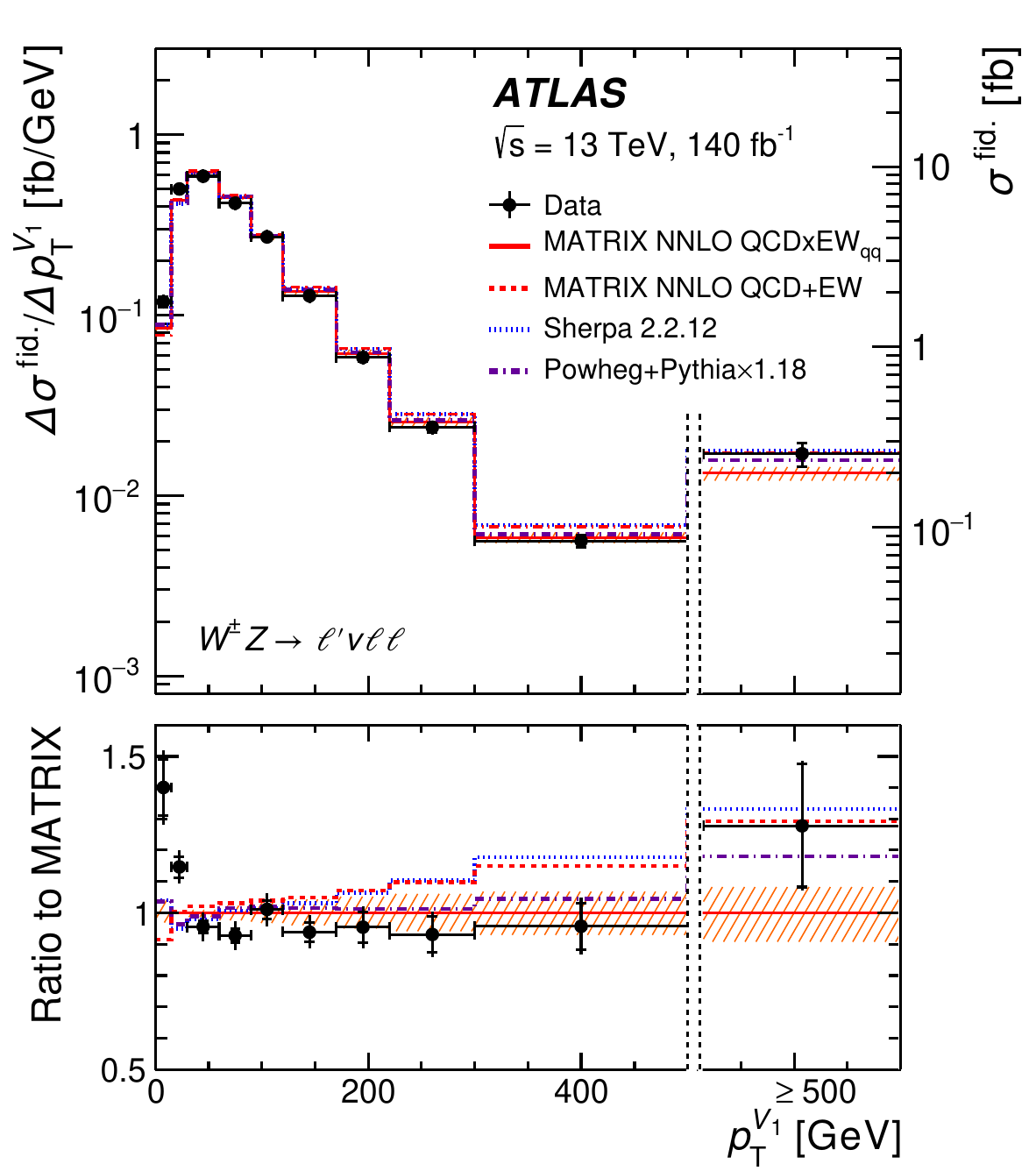}
\caption{The measured differential cross sections of $WZ\to 3\ell1\nu$ in bins of $N_{jets}$ and $p_{\text{T}}^{V1}$~\cite{bib:WZ-FullRun2}.}
\label{fig:WZ-diff}
\end{figure}

$W^{\pm}Z$ productions are measured in tri-lepton decay channel with ATLAS full Run 2 data~\cite{bib:WZ-FullRun2} following up previously published partial Run 2
WZ measurement~\cite{bib:WZ-PartialRun2}, full Run 2 WZ joint polarization measurement~\cite{bib:WZ-JointPolar} and the dedicated high $p_{\text{T}}$ polarization
measurement~\cite{bib:WZ-HighPtPolar}. The new results have shown more precise measurements in total and differential cross sections in comparsion with
NNLO QCD and NLO EWK state-of-art SM predictions. In this new analysis result, more and new kinematic variables are measured differentially as compared to former results such as
$\Delta\phi(\ell^W,Z)$, $p_{\text{T}}^{V1}$ (harder boson), $p_{\text{T}}^{V2}$ (softer boson), $\phi^{\ast}_W$, $\phi^{\ast}_Z$, $p_{\perp}(\Sigma\hat{p}^Z,Z,\ell^W)$, $p_{\perp}(\ell^{+,Z},\ell^{-},\ell^W)$,
while the triple products are sensitive to probe CP-violating SM EFT operators. As shown in Figure~\ref{fig:WZ-diff}: in bins of $p_{\text{T}}^{V1}$, MATRIX $\oplus$ and $\otimes$ schemes
are compared with notable difference at high $p_{\text{T}}$ region; in bins of $N_{\text{jets}}$, Sherpa modeling with more complete description of jet multiplicities is compared and agreed better,
while the low $p_{\text{T}}$ region discrepancies are explainable according to large logarithms in $p_{\text{T}}$/$m_{VV}$.
The measured $W^{\pm}Z$ fiducial cross sections are in agreement with NNLO QCD $\otimes$ EW$_{qq}$ SM prediction from MATRIX using NNPDF3.1nnlo\_as\_118\_luxqed PDF and MSHT20nnlo PDF.
The charge ratio is also measured with $\sim$2\% precision. Both CP-conserving and CP-violating EFT limits are set, with measured $m_{\text{T}}^{WZ}$ distribution and dedicated BDT analysis.

\begin{figure}[htb]
\centering
\includegraphics[width=0.45\textwidth]{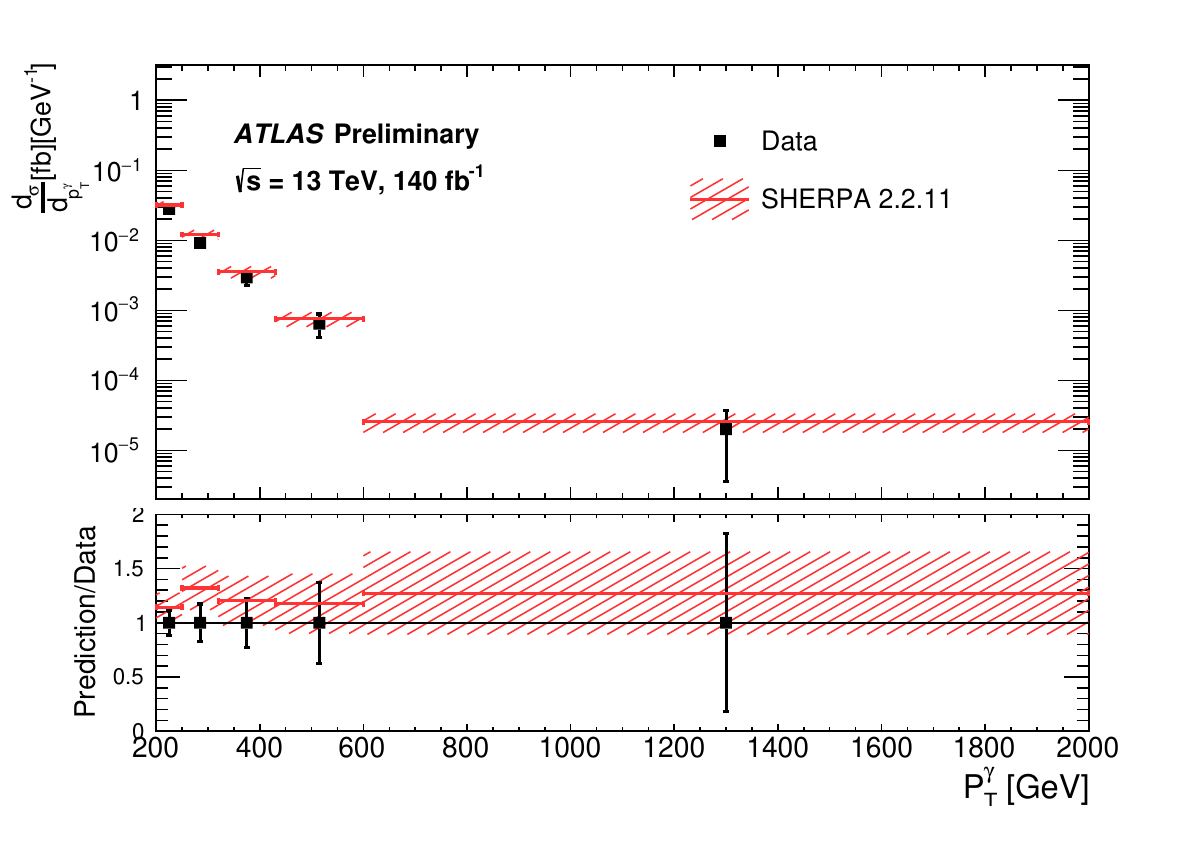}
\includegraphics[width=0.45\textwidth]{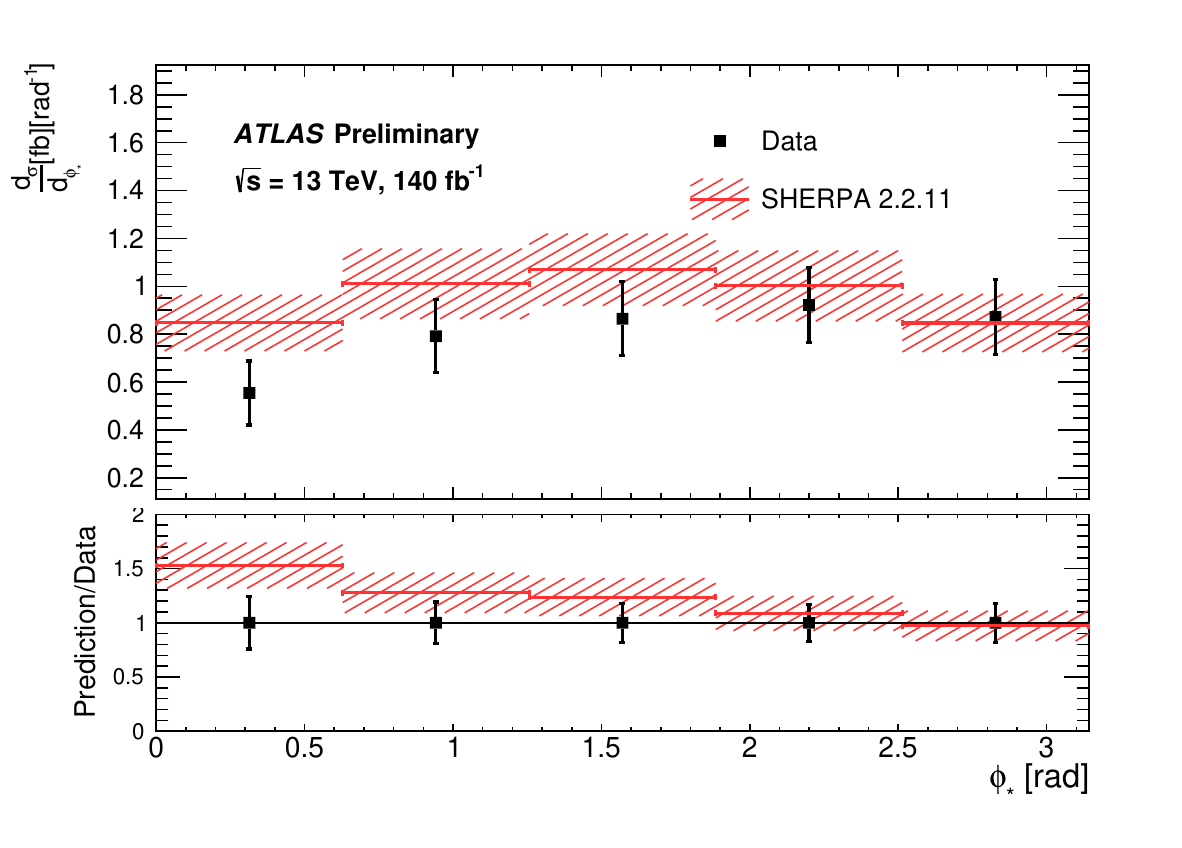}
\caption{The measured differential cross sections of $Z(\ell\ell)\gamma$ in bins of $p_{\text{T}}^{\gamma}$ and $\phi^{\ast}$~\cite{bib:Zy-nTGC}.}
\label{fig:Zy-diff}
\end{figure}

In the very recent differential $Z(\ell\ell)\gamma$ high $p_{\text{T}}$ measurement~\cite{bib:Zy-nTGC}, a derived phasespace is construted on top of the previously published differential measurement
paper~\cite{bib:Zy+jets} by requiring $p_{\text{T}}^{\gamma} >$ 200 GeV, narrow $Z$ mass window and jet veto selections. With the differential cross section measured in bins of $p_{\text{T}}^{\gamma}$,
95\% C.L. limits on the nTGCs are set utilizing the latest SM EFT $SU(2)_L \otimes U(1)_Y$ fully gauge invariant formulation~\cite{bib:nTGC-theory} for the first time at LHC.
Given all previous $Z\gamma$ nTGC limits are using $U(1)$ symmetry treatment leading to an overestimation of the limits on the nTGC vertex form factors by two orders of magnitudes,
such new nTGC results are worth great attention and physical interest to ensure the BSM extension staying close enough to SM symmetry treatment.
A new observable $\phi^{\ast}$, defined as the angle between scattering plane and decay plane of $Z$ in the $\ell^{+}\ell^{-}$ frame to be sensitive to the interferences
between SM and pure BSM terms, is also measured differentially besides the $p_{\text{T}}^{\gamma}$ differential measurement, as shown in Figure~\ref{fig:Zy-diff}.

In conclusion, ATLAS recent results in precisely measuring the EWK diboson productions mark significant precision milestone in understanding SM EWK sector and exmaming the perturbative
QCD and EWK high order predictions/modelings. State-of-Art generators, parton showers and PDF sets are taken into account in the analysis result comparisons providing intensive information
as feedback to SM theory process modeling and BSM interpretations. The rich data statistics from ATLAS Run 2 also enable more precise measurements in charge asymmetries in $W^{+}W^{-}$
and $W^{\pm}Z$ while also offering more sensitive probe to CP-violating signatures and limits sets. Latest SM EFT progress is consulted with $SU(2)_L \otimes U(1)_Y$ fully gauge invariant
nTGC formulation taken into account in $Z(\ell\ell)\gamma$ nTGC analysis leading to significantly relaxed constraints on nTGC form factors compared to previously used $U(1)$ only symmetry
treatment in old formulations.

\end{document}